\documentstyle[preprint,prl,aps,epsf]{revtex}
\def\ba{{\bf a}}

\def\bH{{\bf H}}
\def\bP{{\bf P}}
\def\bz{{\bf \hat{z} }}
\def\cS{{\cal S}}
\def\cC{{\cal C}}
\def\unit{\rm $\times 10^{-6}$cgs/mole}
\begin{document}
\draft
\title{
Novel Magnetic Properties of Carbon Nanotubes
}
\author{Jian Ping Lu}
\address{
Department of Physics and Astronomy,
University of North Carolina at Chapel Hill,
Chapel Hill, North Carolina 27599 \\
jpl@physics.unc.edu
}
\date{Received \today}
\maketitle
\begin{abstract}
An external magnetic field is found to have strong
effects on the electronic structure of carbon nanotubes.
A field-induced metal-insulator transition is predicted
for all pure nanotubes.
In a weak field, nanotubes exhibit
both large diamagnetic and paramagnetic responses which depend on the
field direction, the position of the Fermi energy, the helicity, and
the size of the nanotube.
Universal scalings are found for
the susceptibility as functions of the Fermi energy,
the temperature, and the size of nanotubes.
These results are in agreement with
experiments.
\end{abstract}
\bigskip

\narrowtext

The exciting discovery of carbon nanotubes\cite{iijima}
stimulated a large number of theoretical studies on
their electronic properties.
Both  tight-binding \cite{hamada,white}
and first principles calculations\cite{mintmire}
predicted that nanotubes can be either metallic or
semiconducting depending on their helicity and size.
Several recent experiments demonstrated
some unusual properties.
Giant magnetoresistance and indications of a field-induced
metal-insulator transition are found in transport measurements\cite{langer}.
Large diamagnetic susceptibilities are found for a magnetic
field both perpendicular
and parallel to the tube axis\cite{wang,ramirez}.

The magnetic properties of nanotubes were studied by Ajiki
and Ando\cite{ajiki} using the ${\bf k \cdot p}$ perturbation method
However, they found that the susceptibility $\chi_{\perp}$ (for the
field $\bH$ perpendicular to the tube axis $\bz$) is three orders of magnitude
larger than $\chi_{\parallel}$ (when $\bH \parallel \bz$). This finding
disagrees
with experiments, where it is found that the two
are comparable\cite{wang}.
The ${\bf k \cdot p}$ calculation
is valid only if the Fermi energy is at the center of the band
(half filling), and it only provides information
about the band structure near the Fermi energy.
The orbital magnetism depends on the total band energy which
requires the calculation of the total
band structure. In addition, in
real materials it is likely that
the system is not exactly at the half filling (corresponds to the case
of a finite carrier density in graphite). Thus, a calculation which
includes the field dependence of the complete $\pi$ band is
necessary to understand the magnetic response of the nanotube.

In this letter we report the results of such a calculation
using the tight-binding model and the London approximation.
We obtained the following results.
1) A magnetic field
induced metal-insulator transition is predicted for all pure
nanotubes, the transition depends on the helicity,
the radius $R$, and the magnetic field direction.
2) The weak-field magnetic susceptibility is large
and increases linearly with the size of nanotubes, $\chi \sim R$;
it can be either diamagnetic or paramagnetic
and is sensitive to the position of the Fermi energy $\epsilon_F$.
3) Associate with each nanotube is a unique energy scale
$\Delta_0$; the scaled susceptibility
$\chi/R$ is found to be a universal function of
$\epsilon_F/\Delta_0$ and $k_BT/\Delta_0$
for each family of nanotubes.
4) For typical nanotubes, $\chi \sim -250$\unit, and $|\chi|$ decreases
with increasing $T$. These results are in
agreement with recent experiments\cite{wang,ramirez}.

We use the nearest-neighbor tight-binding hamiltonian
to calculate the band structure formed by the $\pi$ orbitals.
This hamiltonian has been shown to be
an excellent approximation for calculating the electronic structure
of fullerene-related materials such
as the large fullerene molecules\cite{lu}, the solid fullerite and
fullerides\cite{gelfand}, and the nanotubes\cite{hamada}.
Including the effect of a magnetic field in such a model is
straightforward in the London approximation, which
has been used successfully for studying
the ring current and the magnetic response of $C_{60}$ and $C_{70}$
molecules\cite{pasquarello}.
The symmetry of nanotubes have bees studied by several groups\cite{hamada}.
We follow the elegant approach of White et al.\cite{white}.
The structure of a nanotube is defined as the conformal mapping of a
strip of a two-dimensional (2-D) graphitic lattice onto the surface of a
cylinder.
Each nanotube is uniquely characterized by a 2-D
lattice vector ${\bf L}=n_1{\bf a}_1+n_2{\bf a}_2=[n_1,n_2]$,
where ${\bf a_1, a_2}$ are 2-D primitive lattice vectors.
The set of integers $[n_1,n_2]$ determines the geometric
properties of the nanotube. The radius of the tube is
$R=\frac{L}{2\pi}=\frac{\sqrt{3}d_0}{2\pi}\sqrt{n_1^2+n_2^2+n_1n_2}$,
where $d_0$ is the C-C bond length.
There are two symmetry operations: $\cC_N$ and $\cS(\alpha,h)$.
$\cC_N$ is a $N$-fold rotation along the axis, where $N$ is the largest
common denominator of $n_1$ and $n_2$.
The screw translation $\cS(h,\alpha)$ represents a rotation of $\alpha$
about the axis followed by a translation of $h$ along
the axis. The parameters $h$ and $\alpha$ are
determined from a 2-D lattice vector $\bP=p_1\ba_1+p_2\ba_2$,
where $p_1,p_2$ are integers which satisfy the condition
$p_2n_1-p_1n_2=N$\cite{white}.
In the London approximation, the hopping
between site $i$ and site $j$ is modified by a phase factor due to
the presence of a magnetic field,
$V_{ij}=V_0\exp{(i\frac{2\pi}{\phi_0}
\int_{i}^{j} {\bf A(r)} \cdot d {\bf r})}$.
Here $V_0$ is the nearest-neighbor hopping amplitude,
${\bf A(r)}$ is the vector potential associated with the magnetic field,
and $\phi_0=\frac{h c}{e}$ is the flux quantum.

For the special case of a uniform magnetic field
$\bH$ parallel to the tube axis $\bz$, both $\cC_N$ and $\cS(\alpha,h)$
remain
symmetry operations of the tube. In this case the hamiltonian can
be solved analytically and we obtain the following band structure,
\begin{eqnarray}
\epsilon_n(\kappa) &=& \pm V_0 \left[ 3+2\cos (\delta_1)+2\cos (\delta_2)
+2\cos (\delta_1+\delta_2) \right]^{1/2}, \nonumber \\
& &\qquad\qquad\qquad\qquad n=0,1,\cdots,N-1 \ , \nonumber \\
\delta_1 &=& \frac{n_1\kappa-2\pi n p_1}{N} +\beta (n_1+2n_2) \ , \\
\delta_2 &=& \frac{n_2\kappa-2\pi n p_2}{N} -\beta (n_2+2n_1) \ , \nonumber \\
\beta &=& \frac{3d_0^2H}{4\phi_0} \nonumber \ .
\end{eqnarray}
Here $\kappa$ is the pseudo-momentum
associate with the screw translation $\cS(\alpha,h)$. For
$H=0$ we recover the result of White et al.\cite{white}.

For an idea nanotube each $\pi$ orbital
contributes one electron, the Fermi energy is at the center of the band,
$\epsilon_F=0$. From Eq.(1) one finds that,
in addition to the dependence on the helicity and the radius
of the nanotube, the band gap $\Delta=2\min \{ \epsilon_n(\kappa)\}$
varies strongly with the magnetic field.
Our calculations show that $\Delta$ is a simple period function of
the flux, $\phi=\pi R^2H$, with a period of $\phi_0$,
\begin{equation}
\Delta (H)  = \left\{ \begin{array}{ll}
	\Delta_0\frac{3\phi}{\phi_0}\ , & \mbox{~~$0\leq \phi \leq \phi_0/2$} \\
	\Delta_0\left|3-\frac{3\phi}{\phi_0}\right|\ , & \mbox{~~$\phi_0/2 \leq \phi
\leq \phi_0$}
	\end{array} \right.
\end{equation}
for $n_1-n_2 = 3q$, and
\begin{equation}
\Delta (H)  = \left\{ \begin{array}{ll}
	\Delta_0\left|1-\frac{3\phi}{\phi_0}\right|\ ,  & \mbox{~~$0\leq \phi \leq
\phi_0/2$} \\
	\Delta_0\left|2-\frac{3\phi}{\phi_0}\right|\ ,  & \mbox{~~$\phi_0/2 \leq \phi
\leq \phi_0$}
	\end{array} \right.
\end{equation}
for $n_1-n_2 = 3q\pm 1$, where
\begin{equation}
\Delta_0\approx \frac{V_0d_0}{R} \
\end{equation}
is a characteristic energy scale associated with each nanotube.

{}From above equations one can draw several conclusions.
1) In the absence of the magnetic field there are two types
of nanotubes: those with $n_1-n_2 = 3q$ are metallic and
those with $n_1-n_2=3q\pm 1$ are semiconducting.
2) In the presence of a magnetic field along the tube axis a gap
opens up for the metallic tubes. In contrast, the gap decreases
for the semiconducting tubes, reaching zero at one-third of the flux
quantum.
Therefore a field-induced metal-insulator transition is expected for
all nanotubes.
3) The gap scales inversely with the tube
radius. Its dependence on the magnetic field is controlled
by the characteristic field $H^*=\phi_0/(\pi R^2)$, which
can be small for typical nanotubes. For example,
taking $V_0 \sim 2.6$eV, $d_0=1.43$\AA, one obtains
$\Delta \sim 30$meV for a nanotube of radius $R \sim 12$nm.
For such a gap, a magnetic field of $3$ T will
reduce the gap to zero. Such a field is well within the reach of most
experiments.
4) Because of the sensitive dependence of the band structure
on the magnetic field, a
large magnetoresistance is expected for all carbon nanotube.

For a uniform magnetic field making an arbitrary angle $\theta$
with the tube axis, neither $\cC_N$ nor
$\cS(\alpha,h)$ are symmetry operations of the nanotube.
But a translation along the tube axis with the lattice vector
${\bf T}=\frac{2n_1+n2}{N}\ba_1-\frac{n_1+2n_2}{N}\ba_2$
is still a good symmetry operation.
For this case, one can obtain the band structure numerically.
It is found that the gap is no longer a simple periodic function of the
flux. However,
our calculations show that for each family of nanotubes the
reduced gap
$\Delta(H,\theta)/\Delta_0$ is a universal function, when
the magnetic field is scaled by $H^*$.
This universal function is shown in Fig.1 for several field directions and
for both types of nanotubes. The universal scaling relation
enables one to estimate the
gap of any nanotube in an arbitrary magnetic field.
{}From Fig.1 one observes that the effect of field on the band gap is
reduced when the field direction deviates from the tube axis.
This does not imply that physical properties
such as the susceptibility are less affected, as we will show.

The strong field dependence of the band structure
suggests a large magnetic susceptibility $\chi$.
There are two contributions to $\chi$:
the Pauli term $\chi_P$ (due to the electron spin)
and the orbital term
$\chi_{orb}$ (due to the change in the band energy). Because of the low
density of states near the Fermi energy $\chi_{P}$ is negligible\cite{note1},
$\chi=\chi_P+\chi_{orb}\approx \chi_{orb}$.
At $T=0$, $\chi$ can be calculated from
the second derivative of the total energy, but
at a finite temperature the free energy should be used\cite{mcclure},
\begin{eqnarray}
\chi &=& - \left. {\frac{\partial^2 F(H,T)}{\partial H^2}} \right|_{H=0}\ ,
	 \\
F(H,T) &=& - k_BT \sum_{n,\kappa} \ln \left[
	1+\exp \left( {-\frac{\epsilon_n(\kappa,H)-\mu)}{k_BT}} \right)
	\right] \ . \nonumber
\end{eqnarray}
The summation is over the complete
band. It is important to include the effect of the magnetic field
on the complete band structure because
$\chi$ is calculated from the total
energy which includes contributions far away from the Fermi level.
Thus, a perturbative calculation near the Fermi energy
gives an inaccurate result in general

In the case of ideal carbon nanotubes at $T=0$, the Fermi energy is at
the center of the band $\epsilon_F=0$.
It is found that for $\bH \parallel \bz$ the
metallic nanotubes ($n_1-n_2=3q$) are
paramagnetic while the semiconducting nanotubes ($n_1-n_2=3q\pm 1$)
are diamagnetic. In contrast, for
$\bH \parallel \bz$ all nanotubes are diamagnetic (Fig.2 and Fig.3).
The magnitude of the susceptibility increases linearly with the
size of the nanotube regardless whether it is diamagnetic or
paramagnetic, $|\chi| \sim R$.
In general $\chi$ is a sensitive function of
$\theta$.
Empirically it is found that $\chi(\theta)=a+b\cos (2\theta)$
is a good approximation (Fig.2).
This unusual dependence on the field direction suggests
that magnetic poling is a possible method of aligning
nanotubes.

An important result found is that $\chi$ is sensitive
to the position of the Fermi energy. A small change in the carrier density
(hence the Fermi energy) from the half filling,
which is likely in real materials, can
lead to a large change in $\chi$.
The variation of $\chi$ on $\epsilon_F$
depends on the characteristic energy scale $\Delta_0$.
It is found that $\chi/R$ as a function of $\epsilon_f/\Delta_0$ is
universal for each family of nanotubes.
Fig.3 shows examples for both $\bH \parallel \bz$ and $\bH \perp \bz$.
One finds that in the case of a perpendicular field, $\chi$ remains
diamagnetic in the vicinity of the  half filling, while for a
parallel field a small deviation from the half filling
changes the susceptibility dramatically.
(Notice that for the semiconducting nanotubes, the Fermi energy jumps
from $\epsilon_F=0$ to $|\epsilon_F|>0.5\Delta_0$ for a small change
in the carrier density. For a typical nanotube of radius $R\sim 12$nm, the
carrier density at $\epsilon_F=\pm \Delta_0$ is around
$1.0\pm 2.5\times 10^{-5}$.)

Finally, in Fig.4 we show the temperature dependence of $\chi$.
In all cases the magnitude of $\chi$ decreases
with increasing $T$. Again, universal scaling is obtained if
the temperature is scaled by $T^*=\Delta_0/k_B$.
For a typical nanotube of radius $R\sim 12$nm, $\Delta \sim 30$meV,
$T^* \sim 330$K.

Our calculation provides a qualitative explanation for the unusual
transport and magnetic properties observed in recent experiments.
The large magnetoresistance observed in nanotube
bundles\cite{langer} is likely related to the field dependence of the band
structure near the the Fermi level.
If so, our calculations predict that the magnetoresistance
should depend on the field direction and be sensitive to the
carrier density.
The magnetic susceptibility was measured by two
groups, one group for randomly
oriented carbon nanotubes\cite{ramirez} and the other
aligned nanotube bundles\cite{wang}. At low
temperature unusually large diamagnetic susceptibility is found,
$\chi \sim -200$ to $-300$ \unit\ for the magnetic field both
perpendicular and parallel to the tube axis. From this data we
can estimate that the typical radius of nanotubes in those samples
is around $R \sim 7$ to $15$nm, a value close to that reported
by experiments. In addition, both experiments found that $|\chi|$
decreases substantially from low temperature
to room temperature\cite{note2}, suggesting a
characteristic temperature around $300K$.
{}From our calculation, this observation corresponds to
a typical nanotube size of $R \sim 12$nm, consistent with the
above estimate.
One discrepancy remains a puzzle.
Experimentally, $\chi$ is more diamagnetic when the field is parallel to
the tube axis than when it is perpendicular. Our calculation
suggests the reverse in most
cases (Fig.3). Though it is possible to chose a certain $\epsilon_F$ such
that our results agree with experiments,
we feel that existing experiments
are not well enough controlled for such a quantitative comparison.
For example, the issue of uniformity of nanotubes and the
fact that most of them contain multiple shells should be
addressed.
Clearly, further experiments, such as the doping dependence,
are needed for
a quantitative test of the present calculations.

In conclusion, we have shown that novel properties are to be expected for
carbon nanotubes in a magnetic field.
A field-induced metal-insulator transition is predicted
for all pure nanotubes. In general, a large
magnetoresistance is expected due to the sensitive dependence
of the electronic
structure on the external magnetic field.
The weak-field magnetic susceptibility is large and increases with the nanotube
radius. The susceptibility
can be either diamagnetic
or paramagnetic depending sensitively on the helicity
of the nanotube,
the field direction, and the position of the Fermi energy.
A characteristic energy $\Delta_0$ exists for each nanotube.
Universal scaling is found for $\chi$ as a function of
the reduced Fermi energy $\epsilon_F/\Delta_0$ and
the reduced temperature $k_BT/\Delta_0$.
Both the magnitude and the temperature dependence of the
susceptibility are in agreement with
recent experiments. Using the current experimental
data, it is estimated that typical nanotube radii are
around $R \sim 7\ -\ 15$nm, consistent with direct observations.
Our results indicate that careful measurements of the magnetic susceptibility
provides an efficient probe to characterize nanotubes.
The directional dependence of the magnetic response suggests
the magnetic poling as a possible method of aligning nanotubes.
The novel and unusual magnetic properties of nanotubes may
have promising applications in areas such as the magneto-electronics
and the magnetic field detectors.

\acknowledgments
The author thanks O. Zhou for communicating
their experimental results prior to the publication.
This work is supported by The Petroleum Research Fund and
the University of North Carolina at Chapel Hill.


\vbox{
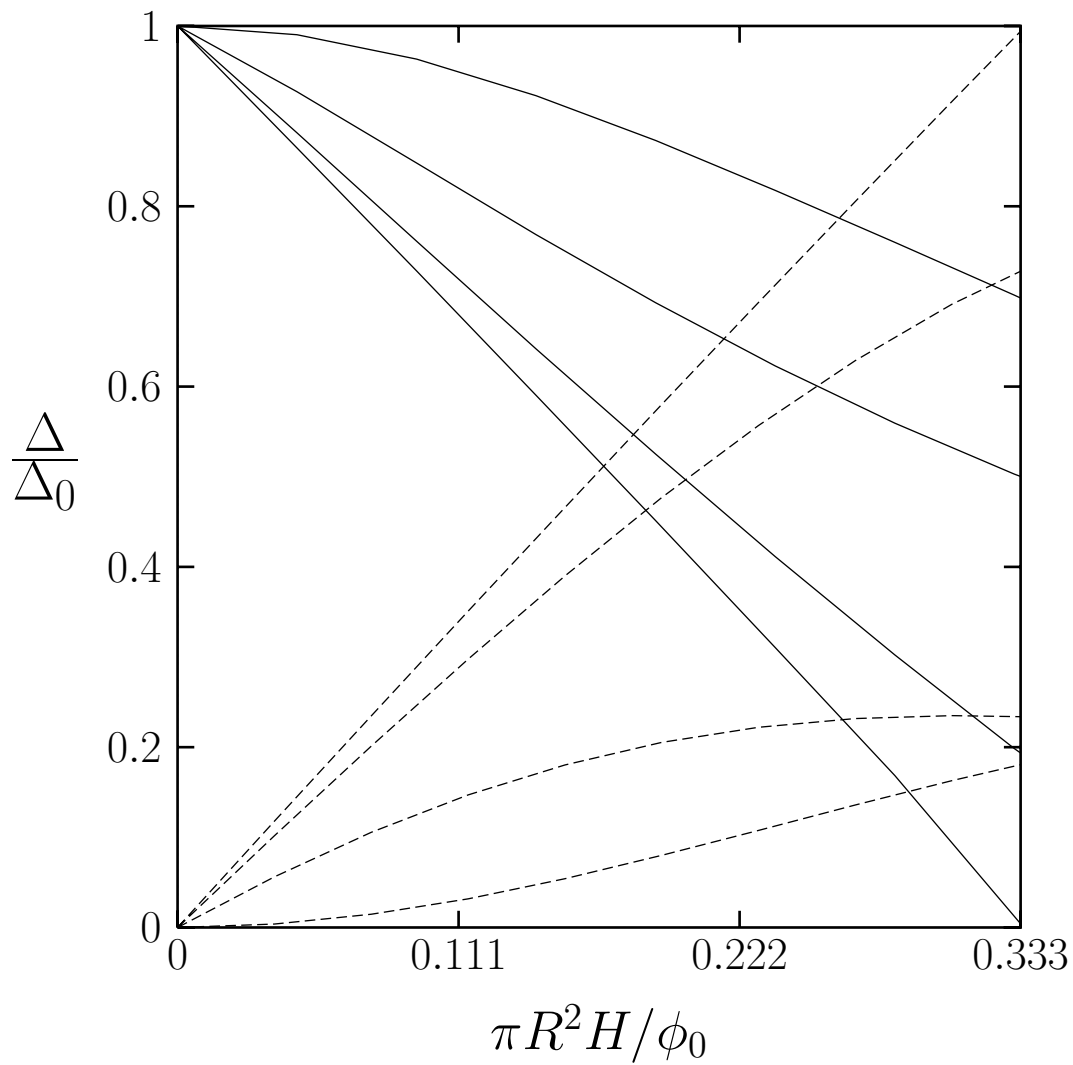
\begin{figure}
\begin{center}
\setlength{\unitlength}{0.1bp}
\begin{picture}(3960,3702)(0,0)
\put(2188,-149){\makebox(0,0){{\LARGE $\pi R^2 H/\phi_0$}}}
\put(100,1951){\makebox(0,0)[b]{\shortstack{{\huge
$\frac{\Delta}{\Delta_0}$}}}}
\put(3777,151){\makebox(0,0){{\Large$0.333$}}}
\put(2718,151){\makebox(0,0){{\Large$0.222$}}}
\put(1659,151){\makebox(0,0){{\Large$0.111$}}}
\put(600,151){\makebox(0,0){{\Large$0$}}}
\put(540,3651){\makebox(0,0)[r]{{\Large$1$}}}
\put(540,2971){\makebox(0,0)[r]{{\Large$0.8$}}}
\put(540,2291){\makebox(0,0)[r]{{\Large$0.6$}}}
\put(540,1611){\makebox(0,0)[r]{{\Large$0.4$}}}
\put(540,931){\makebox(0,0)[r]{{\Large$0.2$}}}
\put(540,251){\makebox(0,0)[r]{{\Large$0$}}}
\end{picture}

\end{center}
\vskip1in
\caption{The reduced gap $\Delta/\Delta_0$ as a function of the
magnetic field $H$ for different
field directions.
Solid lines: semiconducting tubes ($n_1-n_2=3q\pm 1$),
$\theta=\pi /2,\pi /3,\pi/6, 0$ (top down).
Dashed lines: metallic tubes ($n_1-n_2=3q$),
$\theta=\pi /2,\pi /3,\pi/6, 0$ (bottom up).
$\theta$ is the angle between the field direction and the
tube axis, $\Delta_0=V_0d_0/R$.
For all figures presented in this paper we use
$V_0=2.6$eV, $d_0=1.43$\AA.}
\end{figure}
}

\vbox{
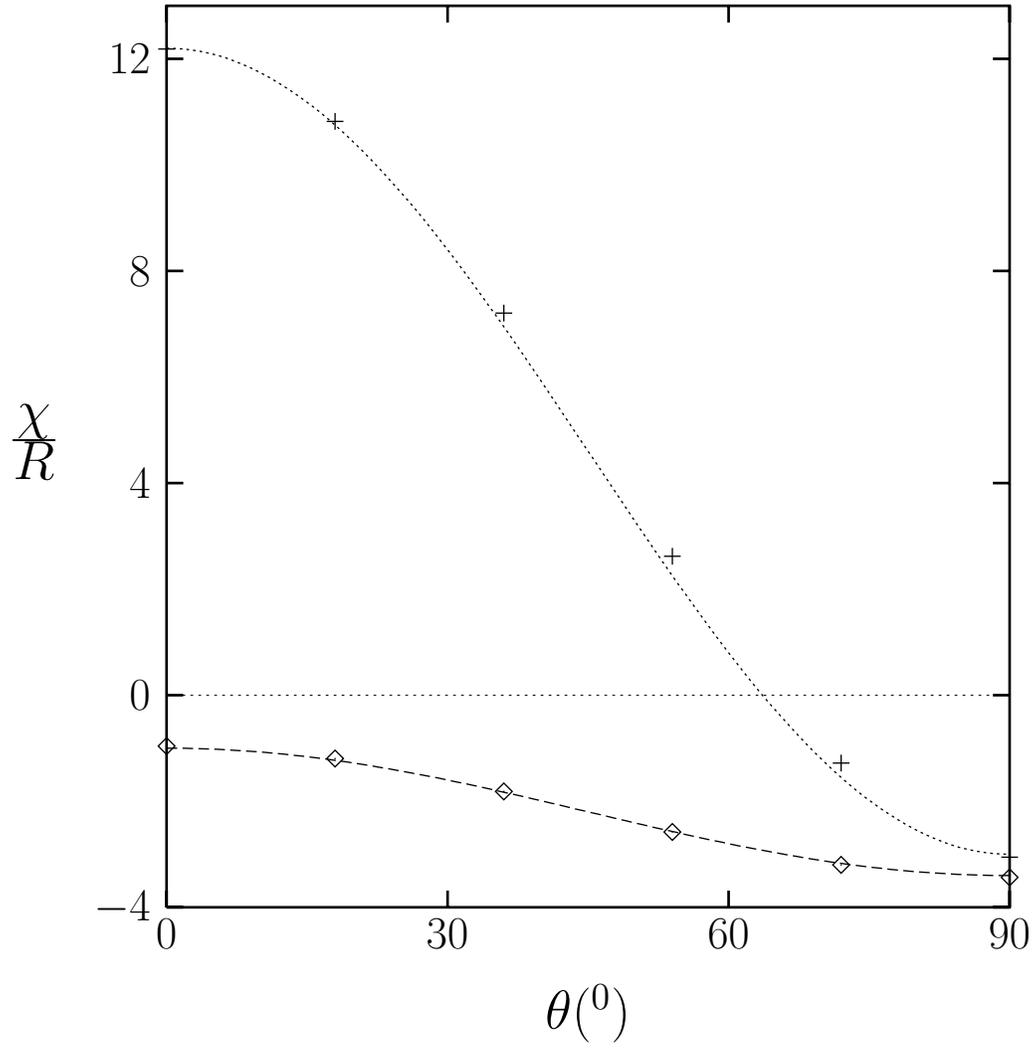
\begin{figure}
\begin{center}
\setlength{\unitlength}{0.1bp}
\begin{picture}(3960,3702)(0,0)
\put(2188,-149){\makebox(0,0){\LARGE $\theta$($^0$)}}
\put(100,1951){\makebox(0,0)[b]{\shortstack{\huge $\frac{\chi}{R}$}}}
\put(3777,151){\makebox(0,0){{\Large$90$}}}
\put(2718,151){\makebox(0,0){{\Large$60$}}}
\put(1659,151){\makebox(0,0){{\Large$30$}}}
\put(600,151){\makebox(0,0){{\Large$0$}}}
\put(540,3451){\makebox(0,0)[r]{{\Large$12$}}}
\put(540,2651){\makebox(0,0)[r]{{\Large$8$}}}
\put(540,1851){\makebox(0,0)[r]{{\Large$4$}}}
\put(540,1051){\makebox(0,0)[r]{{\Large$0$}}}
\put(540,251){\makebox(0,0)[r]{{\Large$-4$}}}
\end{picture}

\end{center}
\vskip1in
\caption{The scaled susceptibility $\chi/R$
as a functions of the
angle $\theta$ between the magnetic field direction and the
tube axis. $\epsilon_F=0,\ T=0$.
Plus: $n_1-n_2=3$. Diamonds: $n_1-n_2=3q+1$. Lines are fits
to the functional form $a+b\cos(2\theta)$.
The case of $n_1-n_2=3q-1$ is very similar to that of
$n_1-n_2=3q+1$. $\chi$ is in units of
$10^{-6}$cgs/mole, $R$ is in units of \AA.}
\end{figure}
}

\vbox{
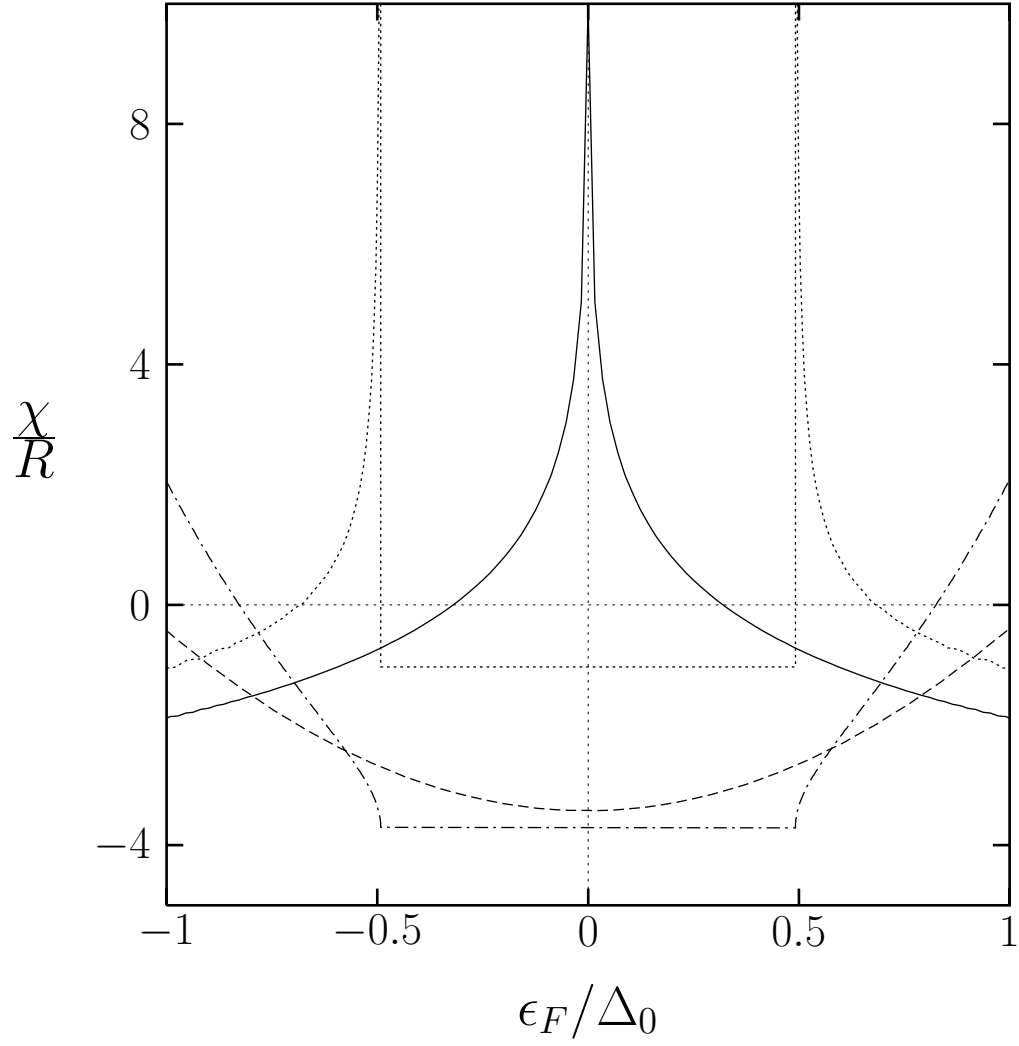
\begin{figure}
\begin{center}
\setlength{\unitlength}{0.1bp}
\begin{picture}(3960,3702)(0,0)
\put(2188,-149){\makebox(0,0){\LARGE $\epsilon_F/\Delta_0$}}
\put(100,1951){\makebox(0,0)[b]{\shortstack{\huge $\frac{\chi}{R}$}}}
\put(3777,151){\makebox(0,0){{\Large$1$}}}
\put(2983,151){\makebox(0,0){{\Large$0.5$}}}
\put(2189,151){\makebox(0,0){{\Large$0$}}}
\put(1394,151){\makebox(0,0){{\Large$-0.5$}}}
\put(600,151){\makebox(0,0){{\Large$-1$}}}
\put(540,3198){\makebox(0,0)[r]{{\Large$8$}}}
\put(540,2291){\makebox(0,0)[r]{{\Large$4$}}}
\put(540,1384){\makebox(0,0)[r]{{\Large$0$}}}
\put(540,478){\makebox(0,0)[r]{{\Large$-4$}}}
\end{picture}

\end{center}
\vskip1in
\caption{Universal scaling of $\chi/R$ as a function of
the reduced Fermi energy $\epsilon_F/\Delta_0$. $T=0$.
Solid line: $\bH \parallel \bz, n_1-n_2=3q$.
Dotted line: $\bH \parallel \bz, n_1-n_2=3q+1$.
Dashed line: $\bH \perp \bz, n_1-n_2=3q$.
Dot-Dashed line: $\bH \perp \bz, n_1-n_2=3q+1$.
The case of $n_1-n_2=3q-1$ is very similar to that of
$n_1-n_2=3q+1$. $\chi/R$ is in units of $10^{-6}$cgs/mole/\AA.}
\end{figure}
}

\vbox{
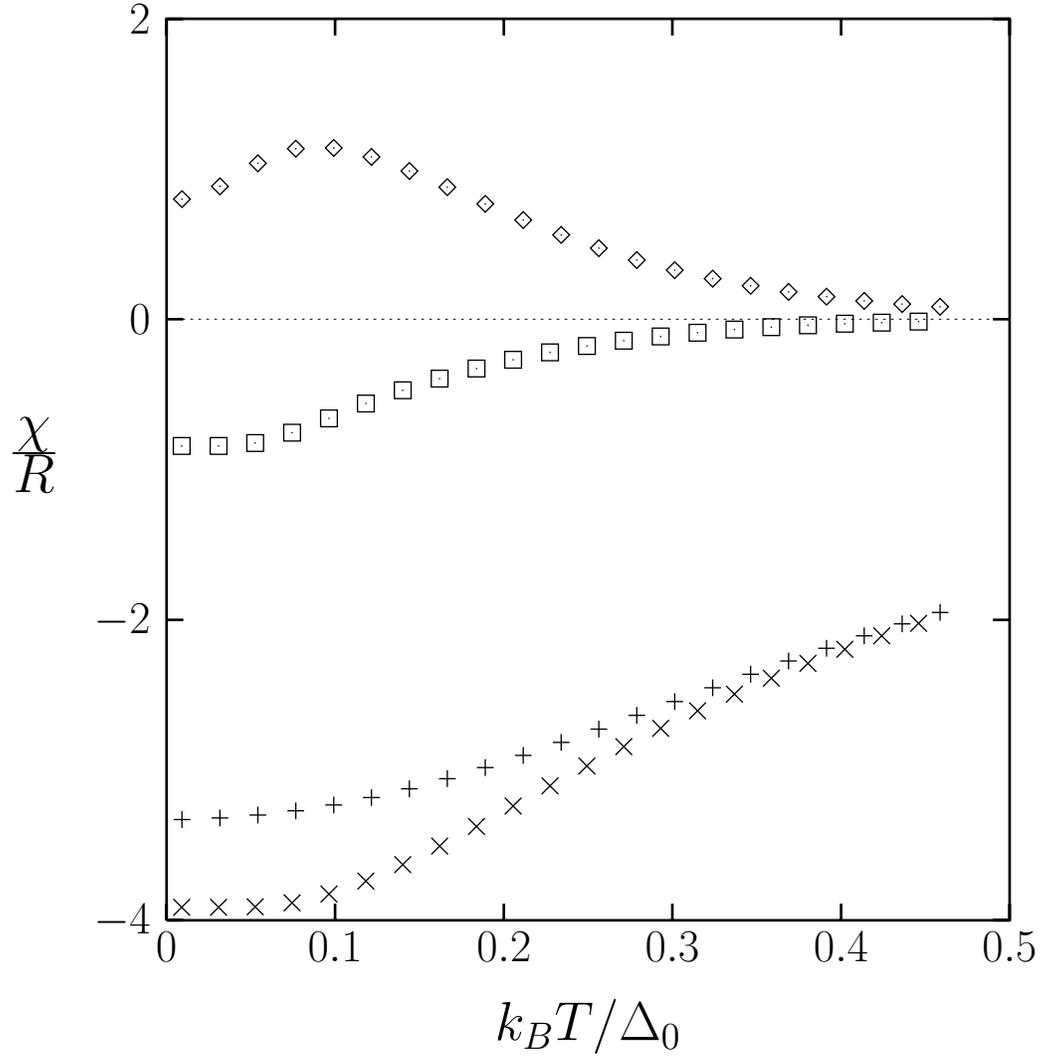
\begin{figure}
\begin{center}
\setlength{\unitlength}{0.1bp}
\begin{picture}(3960,3702)(0,0)
\put(2188,-149){\makebox(0,0){\LARGE $k_BT/\Delta_0$}}
\put(100,1951){\makebox(0,0)[b]{\shortstack{\huge $\frac{\chi}{R}$}}}
\put(3777,151){\makebox(0,0){{\Large$0.5$}}}
\put(3142,151){\makebox(0,0){{\Large$0.4$}}}
\put(2506,151){\makebox(0,0){{\Large$0.3$}}}
\put(1871,151){\makebox(0,0){{\Large$0.2$}}}
\put(1235,151){\makebox(0,0){{\Large$0.1$}}}
\put(600,151){\makebox(0,0){{\Large$0$}}}
\put(540,3651){\makebox(0,0)[r]{{\Large$2$}}}
\put(540,2518){\makebox(0,0)[r]{{\Large$0$}}}
\put(540,1384){\makebox(0,0)[r]{{\Large$-2$}}}
\put(540,251){\makebox(0,0)[r]{{\Large$-4$}}}
\end{picture}

\end{center}
\vskip1in
\caption{The universal dependence of the scaled susceptibility $\chi/R$
on the reduced temperature $k_BT/\Delta_0$. The Fermi energy is at
$\epsilon_F=0.2\Delta_0$.
Squares: $\bH \parallel \bz, n_1-n_2=3q$.
Diamonds: $\bH \parallel \bz, n_1-n_2=3q-1$.
Plus: $\bH \perp \bz, n_1-n_2=3q$.
Crosses: $\bH \perp \bz, n_1-n_2=3q-1$.
The case of $n_1-n_2=3q+1$ is very similar to that of
$n_1-n_2=3q-1$. $\chi/R$ is in units of $10^{-6}$cgs/mole/\AA.}
\end{figure}
}


\begin{references}
\bibitem{note1}
$\chi_P$ measures the density of states at
the Fermi level, $\chi_P \sim 1/R$. For $R \sim 10$nm, $\chi_P$
is three orders of magnitude smaller than $\chi_{orb}$.
\bibitem{note2}
Part of the residual susceptibility observed at the
room temperature may be related to the sample quality, as its
value is close to the susceptibility of the polycrystalline graphite.
This factor may also affect the relative magnitude of $\chi_{\parallel}$
and $\chi_{\perp}$.
\end{references}
\end{document}